
\documentstyle{amsppt}
\magnification=1200
\TagsOnRight
\baselineskip 15pt
\parskip 4pt

\define\Z{Zamolodchikov}
\define\R{\Cal R}

\hfill TURKU-FL-P1, Sept. 17, 1992

\topmatter
\title Some constant solutions to
Zamolodchikov's tetrahedron equations
\endtitle

\author Jarmo Hietarinta \endauthor
\affil Department of Physics,
University of Turku, 20500 Turku, Finland\\
\endaffil

\abstract
In this letter we present constant solutions to the tetrahedron
equations proposed by \Z. In general, from a given solution of the
Yang-Baxter equation there are two ways to construct solutions to the
tetrahedron equation. There are also other kinds of solutions. We
present some two-dimensional solutions that were obtained by directly
solving the equations using either an upper triangular or \Z 's
ansatz.
\endabstract

\endtopmatter

\document
\heading 1. Introduction \endheading
The theory of integrable dynamical systems in 1+1 dimensions (both
continuous PDE and discrete lattice systems) is now fairly well
understood, and at present an increasing amount of research is focused
on generalizations to higher dimensions.

In 1+1 dimensions the key equation for integrability of lattice
systems [1] and quantum inverse transformation [2] is the quantum
Yang-Baxter equation (YBE) [3]
$$
R_{j_1j_2}^{k_1k_2}(u)R_{k_1j_3}^{l_1k_3}(u+v)R_{k_2k_3}^{l_2l_3}(v)=
R_{j_2j_3}^{k_2k_3}(v)R_{j_1k_3}^{k_1l_3}(u+v)R_{k_1k_2}^{l_1l_2}(u).
\tag 1
$$
(Here and in the following summation over repeated indices is understood.)
In dimension $N$ there are $N^6$ equations for $N^4$ variables.

The constant, spectral parameter independent version of (1) is
$$
R_{j_1j_2}^{k_1k_2}R_{k_1j_3}^{l_1k_3}R_{k_2k_3}^{l_2l_3}=
R_{j_2j_3}^{k_2k_3}R_{j_1k_3}^{k_1l_3}R_{k_1k_2}^{l_1l_2}.
\tag2a
$$
This is obtained as the limit $u=v=0$ or $u=v=\pm\infty$, but appears
also independently in the study of quantum groups [4] and knot theory
[5].  In the shorthand notation one writes out only the labels of
subspaces on which the matrices act:
$$
R_{12}R_{13}R_{23}=R_{23}R_{13}R_{12}.\tag 2b
$$

In [6] \Z\ proposed a three dimensional generalization of YBE based on
scattering of straight strings. A lattice interpretation was later
given by Bazhanov and Stroganov [7] and by Baxter [8]. The work [6-8]
produced a spectral parameter dependent equation (with three "spectral
angles" related by rules of spherical trigonometry). As for YBE one
has also a spectral parameter independent formulation, which can be
taken as a special limit or may be considered in its own right.  We
will discuss here only this constant form of \Z 's tetrahedron
equation (ZTE), which is
$$
\R_{j_1j_2j_3}^{k_1k_2k_3}\R_{k_1j_4j_5}^{l_1k_4k_5}
\R_{k_2k_4j_6}^{l_2l_4k_6}\R_{k_3k_5k_6}^{l_3l_5l_6}=
\R_{j_3j_5j_6}^{k_3k_5k_6}\R_{j_2j_4k_6}^{k_2k_4l_6}
\R_{j_1k_4k_5}^{k_1l_4l_5}\R_{k_1k_2k_3}^{l_1l_2l_3}\tag 3a
$$
or in the shorthand notation
$$
\R_{123}\R_{145}\R_{246}\R_{356}=\R_{356}\R_{246}\R_{145}\R_{123}.\tag 3b
$$
Now there are $N^{12}$ equations for $N^6$ variables; even for the
simplest nontrivial case of $N=2$ there are 4096 equations for 64
variables.

There is a natural method of generalizing this to higher dimensions,
see [7,9]. The hierarchy so obtained is not just formal: in the
following we show how each solution of a higher level equation yields
solutions for the lower level, and how each lower level solution can
be used to get solutions at higher level.

There is also another generalization to three dimensions by Frenkel
and Moore [10]:
$$
\Cal F_{123}\Cal F_{124}\Cal F_{134}\Cal F_{234}=
\Cal F_{234}\Cal F_{134}\Cal F_{124}\Cal F_{123}.\tag 4
$$
Here there are only four different subspaces on which the matrices
act. Carter and Saito have shown [11] how both of these fit into a
generalized sequence of higher dimensional equations, ZTE and
Frenkel-Moore equation are just located on different rays starting
from YBE. In this letter we consider only ZTE.

\heading 2. General Aspects \endheading

\subheading{2.1 Symmetries}
Before discussing the solutions it is useful to recall the symmetries
of the tetrahedron equations. This is important for classifying the
solutions, for there is no point in repeating a solution in a form
that can be obtained by one of the allowed transformations.

The symmetries of (3) are basically the same as those of YBE.
First of all there is the invariance under continuous transformations
$$
\R\to \kappa(Q\otimes Q\otimes Q) \R (Q\otimes Q\otimes Q)^{-1},\tag 5
$$
where $Q$ is a nonsingular $N\times N$ matrix and $\kappa$ a nonzero
number.

There are also discrete symmetries:
$$
\align
&\R_{ijk}^{lmn}\to \R_{lmn}^{ijk},\tag 6a\\
&\R_{ijk}^{lmn}\to \R_{i+s,j+s,k+s}^{l+s,m+s,n+s},
\text{ (indices mod $N$) }\tag 6b\\
&\R_{ijk}^{lmn}\to \R_{kji}^{nml}.\tag 6c
\endalign
$$
In writing out the triple indiced object we use the usual matrix
notation and the connection is
$\R_{i+(j-1)N+(k-1)N^2}^{l+(m-1)N+(n-1)N^2}=\R_{ijk}^{lmn}$.  In this
notation (6a) corresponds to the usual matrix transposition. For
$N=2,\,s=1$ (6b) followed by (6a) corresponds to transposition across
the auxiliary diagonal; transformation (6c) corresponds to the
simultaneous exchange of the following columns (and rows): $2
\leftrightarrow 5$, and $4 \leftrightarrow 7$.

\subheading{2.2 From ZTE to YBE}
In the following we need a notation for traces of multi-index
matrices, we use a square bracket to denote the location of the traced
index, e.g. $(\R[2])_{ij}^{kl}=\sum_m \R_{imj}^{kml}$.

Let us now assume that we have a nonsingular solution of (3).
Multiplying (3a) by $(\R^{-1})_{l_1l_2l_3}^{j_1j_2j_3}$ and summing
over the repeated indices yields (2) for $\R[1]$ with renumbering of
indices.  The same result is obtained for $\R[3]$ with
$(\R^{-1})_{l_3l_5l_6}^{j_3j_5j_6}$.  One can obtain an even stronger
result. For that we need the following
\proclaim{Definition} Three double indiced matrices $(A,M,B)$ form an
{\bf associated triple} of Yang-Baxter matrices (AT) if the following
equations hold
$$
\aligned
A_{12}A_{13}A_{23}&=A_{23}A_{13}A_{12},\\
M_{12}M_{13}A_{23}=A_{23}M_{13}M_{12},&\quad
B_{12}M_{13}M_{23}=M_{23}M_{13}B_{12},\\
B_{12}B_{13}B_{23}&=B_{23}B_{13}B_{12}.
\endaligned\tag7
$$
\endproclaim

Trivially $(R,R,R)$ is an AT if $R$ is a constant solution of (2), but
more interestingly, if $R(x)$ is a solution of (1), then
$(R(0),R(x),R(0))$ is also an AT, as can be readily seen by
substituting $u=0$ and/or $v=0$ into (1). This does not exhaust the
solutions, for example $(P,M,P)$, where $P$ is the permutation matrix,
is an AT for an arbitrary $M$.

Let us now return to ZTE. Above we showed that if $\R$ is nonsingular
then $\R[1]$ and $\R[3]$ satisfy (2).  If also $\R^{t_1}$ and
$\R^{t_3}$ (transpose on the first and third index, respectively) are
nonsingular, then after multiplying (3a) by
$((\R^{t_1})^{-1})_{j_1l_4l_5}^{l_1j_4j_5}$ and by
$((\R^{t_3})^{-1})_{l_2l_4j_6}^{j_2j_4l_6}$ one finds that
$(\R[1],\R[2],\R[3])$ must be an AT. This fact can be used in limiting
the ansatz for ZTE.

If the same method is applied to (2) one gets condition of
commutativity for its trace matrices. For constant solutions the only
nontrivial condition is that $R_{im}^{jm}$ and $R_{mi}^{mj}$ commute.
(It should be noted here that nonsingularity is a necessary
requirement and there is a singular solution ($R_{H1.5}$ of [13]) for
which these trace matrices do not commute.)

An interesting open problem is to see if the above works with spectral
parameters. At each level the spectral parameters live in different
spaces, but there is probably a natural projection which is necessary
for obtaining the correct equation for the trace matrices.

\subheading{2.3 From YBE to ZTE by squaring the dimension}
Above we showed how each nonsingular solution of (3) in dimension $N$
yields an AT of the same dimension. We will now show how each such AT
yields a solution of (3) of dimension $N^2$. This is a generalization
of [12,11] where the AT $(R,R,R)$ was used.

Let
$$
\R_{j_1j_2j_3}^{l_1l_2l_3}\equiv
\R_{(j_1^aj_1^b)(j_2^aj_2^b)(j_3^aj_3^b)}
^{(l_1^al_1^b)(l_2^al_2^b)(l_3^al_3^b)}:=
A_{j_1^aj_2^a}^{l_1^al_2^a}
M_{j_1^bj_3^a}^{l_1^bl_3^a}
B_{j_2^bj_3^b}^{l_2^bl_3^b},\tag 8
$$
then we have
$$
\align
\R_{j_1j_2j_3}^{k_1k_2k_3}&\R_{k_1j_4j_5}^{l_1k_4k_5}
\R_{k_2k_4j_6}^{l_2l_4k_6}\R_{k_3k_5k_6}^{l_3l_5l_6}
\\
&=
(A_{j_1^aj_2^a}^{k_1^ak_2^a}
M_{j_1^bj_3^a}^{k_1^bk_3^a}
B_{j_2^bj_3^b}^{k_2^bk_3^b})
(A_{k_1^aj_4^a}^{l_1^ak_4^a}
M_{k_1^bj_5^a}^{l_1^bk_5^a}
B_{j_4^bj_5^b}^{k_4^bk_5^b})
(A_{k_2^ak_4^a}^{l_2^al_4^a}
M_{k_2^bj_6^a}^{l_2^bk_6^a}
B_{k_4^bj_6^b}^{l_4^bk_6^b})
(A_{k_3^ak_5^a}^{l_3^al_5^a}
M_{k_3^bk_6^a}^{l_3^bl_6^a}
B_{k_5^bk_6^b}^{l_5^bl_6^b})\\
&=
[A_{j_1^aj_2^a}^{k_1^ak_2^a}
A_{k_1^aj_4^a}^{l_1^ak_4^a}
A_{k_2^ak_4^a}^{l_2^al_4^a}]
[M_{j_1^bj_3^a}^{k_1^bk_3^a}
M_{k_1^bj_5^a}^{l_1^bk_5^a}
A_{k_3^ak_5^a}^{l_3^al_5^a}]
[B_{j_2^bj_3^b}^{k_2^bk_3^b}
M_{k_2^bj_6^a}^{l_2^bk_6^a}
M_{k_3^bk_6^a}^{l_3^bl_6^a}]
[B_{j_4^bj_5^b}^{k_4^bk_5^b}
B_{k_4^bj_6^b}^{l_4^bk_6^b}
B_{k_5^bk_6^b}^{l_5^bl_6^b}]\\
&=
 [A_{j_2^aj_4^a}^{k_2^ak_4^a}
 A_{j_1^ak_4^a}^{k_1^al_4^a}
 A_{k_1^ak_2^a}^{l_1^al_2^a}]
 [A_{j_3^aj_5^a}^{k_3^ak_5^a}
 M_{j_1^bk_3^a}^{k_1^bl_3^a}
 M_{k_1^bk_5^a}^{l_1^bl_5^a}]
 [M_{j_3^bj_6^a}^{k_3^bk_6^a}
 M_{j_2^bk_6^a}^{k_2^bl_6^a}
 B_{k_2^bk_3^b}^{l_2^bl_3^b}]
 [B_{j_5^bj_6^b}^{k_5^bk_6^b}
 B_{j_4^bk_6^b}^{k_4^bl_6^b}
 B_{k_4^bk_5^b}^{l_4^bl_5^b}]\\
&=
 (A_{j_3^aj_5^a}^{k_3^ak_5^a}
 M_{j_3^bj_6^a}^{k_3^bk_6^a}
 B_{j_5^bj_6^b}^{k_5^bk_6^b})
(A_{j_2^aj_4^a}^{k_2^ak_4^a}
 M_{j_2^bk_6^a}^{k_2^bl_6^a}
 B_{j_4^bk_6^b}^{k_4^bl_6^b})
( A_{j_1^ak_4^a}^{k_1^al_4^a}
 M_{k_1^bk_5^a}^{l_1^bl_5^a}
 B_{k_4^bk_5^b}^{l_4^bl_5^b})
( A_{k_1^ak_2^a}^{l_1^al_2^a}
 M_{j_1^bk_3^a}^{k_1^bl_3^a}
 B_{k_2^bk_3^b}^{l_2^bl_3^b})\\
&=
\R_{j_3j_5j_6}^{k_3k_5k_6}\R_{j_2j_4k_6}^{k_2k_4j_6}
\R_{j_1k_4k_5}^{k_1l_4l_5}\R_{k_1k_2k_3}^{l_1l_2l_3},\tag 9
\endalign
$$
i.e. $\R$ satisfies (3). The third equality sign follows from the
assumption that $(A,M,B)$ forms an AT.  Note that if we use the AT
$(R(0),R(x),R(0))$ the spectral parameter of YBE plays now the role of
an ordinary parameter in a constant solution of (3).

\subheading{2.4 From YBE to ZTE with same dimension}
In the above construction the dimension of the space gets squared.
There is also a method of getting solutions of the same dimension.

Let us assume that $\R_{123}=R_{12}\otimes m_3$. Then it is easy to
see that $\R$ solves (3) if any of the following holds
\item{a)} $m^2=0$: For example $R=A\otimes \pmatrix0&1\\0&0\endpmatrix$
is a solution for arbitrary $A$.
\item{b)} $(m\otimes m)R=R(m\otimes m)=0$: Here we may nevertheless
assume that $m^2\neq0$, thus for a nontrivial solution there is
precisely one nonzero entry on the diagonal. A typical solution is
$$
R=\pmatrix
0 & 0 & 0 & 0\\
0 & a & b & c\\
0 & d & e & f\\
0 & g & h & j\endpmatrix
\otimes\pmatrix1 & 0\\0 & 0\endpmatrix
$$
\item{c)} $R$ is a solution of (2) and $ [R,m\otimes m]=0$.
The last commutation condition is trivially true if $m=1_{2\times2}$,
but depending on $R$ it can have much more general solutions and
provide nontrivial interaction.

The above works equally well for $\R_{123}=m_1\otimes R_{23}$.

\heading 3. Particular Solutions \endheading
We have also searched directly for particular two-dimensional
solutions that do not fit into the general form of Sec. 2.4.  Since
the full set has 4096 quadratic equations in 64 variables it is not
yet feasible to find the complete solution as in [13].  We will now
describe the results of some ansatze that worked well for the
Yang-Baxter equation.  The computations were done using the symbolic
algebra language REDUCE 3.4 [14], the equations were analysed using
the GROEBNER package [15] written for REDUCE.

\subheading{3.1 Upper triangular ansatz}
If the entries on the diagonal are $=1$, then the only new solution is
$$\R_1=
\pmatrix
1 & q & p & k & d & c & b & a\\
0 & 1 &  0 &  p & 0 &  d & 0 &  b\\
0 & 0 &  1 &  q & 0 &  0 &  d & c\\
0 & 0 &  0 &  1 &  0 &  0 &  0 &  d\\
0 & 0 &  0 &  0 &  1 &  q & p & k\\
0 & 0 &  0 &  0 &  0 &  1 &  0 &  p\\
0 & 0 &  0 &  0 &  0 &  0 &  1 &  q\\
0 & 0 &  0 &  0 &  0 &  0 &  0 &  1
\endpmatrix\tag 10
$$
This is clearly a tetrahedron generalization of the Yang-Baxter
solution $R_{H2.3}$ of [13].

We have also looked at solutions with arbitrary nonzero entires on the
diagonal.  This search is still open but so far we have not found
anything interesting.

\subheading{3.2 Bi-diagonal ansatz}
In [6] \Z\ proposed an ansatz for obtaining a spectral parameter
dependent solution. As shown in [9] it amounts to allowing nonzero
entries only on the diagonal and on the auxiliary diagonal with
certain symmetry relations. We take this bi-diagonal without any
additional relations, i.e.
$$
\R_B=\pmatrix
a_1 & 0 & 0 & 0 & 0 & 0 & 0 & a_8 \\
0 & b_2 & 0 & 0 & 0 & 0 & b_7 & 0 \\
0 & 0 & c_3 & 0 & 0 & c_6 & 0 & 0 \\
0 & 0 & 0 & d_4 & d_5 & 0 & 0 & 0 \\
0 & 0 & 0 & e_4 & e_5 & 0 & 0 & 0 \\
0 & 0 & f_3 & 0 & 0 & f_6 & 0 & 0 \\
0 & g_2 & 0 & 0 & 0 & 0 & g_7 & 0 \\
h_1 & 0 & 0 & 0 & 0 & 0 & 0 & h_8
\endpmatrix\tag 11a
$$
To save space we write out only the two diagonals:
$$
\R_B=[a_1,b_2,c_3,d_4,e_5,f_6,g_7,h_8;
a_8,b_7,c_6,d_5,e_4,f_3,g_2,h_1].\tag 11b
$$

First of all we have a purely diagonal solution
$$
\R_2=[a_1,b_2,c_3,d_4,e_5,f_6,g_7,h_8;
0,0,0,0,0,0,0,0],\tag 12
$$
as we have for YBE ($R_{H3.1}$ of [13]).

If there are nonzero entries only on the auxiliary diagonal,
there must already be relations among them, as follows
$$
\R_3=[0,0,0,0,0,0,0,0;
a_8,b_7,c_6,d_5,d_5,c_6,b_7,a_8],\tag 13
$$
c.f. $R_{H1.4}$ of [13].

We have also searched for all solutions where all entries on both
diagonals are actually nonzero and for which the determinant of the
matrix is also nonzero. This resulted in the following solutions:

$$
\R_4=[1,1,\epsilon_1,\epsilon_2,1,\epsilon_1\epsilon_2,
\epsilon_2,\epsilon_2;
q,q,\epsilon_2q,\epsilon_1q,q,\epsilon_1\epsilon_2q,
\epsilon_1q,\epsilon_1q],\tag 14
$$
and
$$
\R_5=
[1,\epsilon_1,\epsilon_2,\epsilon_3,\epsilon_1,
\epsilon_1\epsilon_2\epsilon_3,\epsilon_3,\epsilon_1\epsilon_3;
1,-\epsilon_2\epsilon_3,-\epsilon_2,\epsilon_1\epsilon_2,
-\epsilon_2\epsilon_3,\epsilon_1\epsilon_2\epsilon_3,
\epsilon_1\epsilon_2,-\epsilon_1\epsilon_3].\tag 15
$$ Here $q$ is a free parameter and $\epsilon_i=\pm1$, independently.
Note however, that if all $\epsilon_i=+1$ then the solutions can in
fact be diagonalized.

Within this ansatz there are numerous other solutions with some
entries zero, let me just mention one: $$
\R_6=
[1,\xi^2,\xi,\epsilon_1,\xi^2,\xi\epsilon_1,\epsilon_2,
\xi^2\epsilon_2;
1,0,0,0,0,0,0,0],\tag 16
$$
where $\xi^6=1$.

\subheading{3.3 Other solutions}
We have also searched for solutions for which $\R_{ijk}^{lmn}\neq0$
only if $i+j+k-l-m-n=0$, but this ansatz did not yield any interesting
nonsingular solutions. One nice singular solution is $$
\R_7=
 \delta_{j_1}^{l_1}\delta_{j_2}^{l_2}\delta_{j_3}^{l_3}
-\delta_{j_1}^{l_3}\delta_{j_2}^{l_2}\delta_{j_3}^{l_1}
+\delta_{j_1}^{l_3}\delta_{j_2}^{l_1}\delta_{j_3}^{l_2}.\tag 17
$$

There is a huge number singular solutions,
here are just two that are "arrow-like":
$$
\R_8=\pmatrix
0 & 0 & 0 & a_4 & 0 & a_6 & a_4 & a_8 \\
0 & 0 & 0 & 0 & 0 & 0 & 0 & -a_6 \\
0 & 0 & 0 & 0 & 0 & 0 & 0 & -a_4 \\
0 & 0 & 0 & 0 & 0 & 0 & 0 & 0 \\
0 & 0 & 0 & 0 & 0 & 0 & 0 & -a_6 \\
0 & 0 & 0 & 0 & 0 & 0 & 0 & 0 \\
0 & 0 & 0 & 0 & 0 & 0 & 0 & 0 \\
1 & 0 & 0 & 0 & 0 & 0 & 0 & 0
\endpmatrix,\tag 18
$$
and if the lower left hand corner is zero, then the other entries are free:
$$
\R_9=\pmatrix
0 & 0 & 0 & a_4 & 0 & a_6 & a_4 & a_8 \\
0 & 0 & 0 & 0 & 0 & 0 & 0 & b_8 \\
0 & 0 & 0 & 0 & 0 & 0 & 0 & c_8 \\
0 & 0 & 0 & 0 & 0 & 0 & 0 & 0 \\
0 & 0 & 0 & 0 & 0 & 0 & 0 & e_8 \\
0 & 0 & 0 & 0 & 0 & 0 & 0 & 0 \\
0 & 0 & 0 & 0 & 0 & 0 & 0 & 0 \\
0 & 0 & 0 & 0 & 0 & 0 & 0 & 0
\endpmatrix.\tag 19
$$

\heading 4. Conclusions \endheading
In this latter we have shown that, contrary to the popular pessimistic
view, ZTE does indeed have many solutions. Some of them are inherited
from the YBE but even then they have some extra structure. As shown
above, there are also genuinely new solutions with no such connection.
(The algebraic aspects of these solutions will be discussed elsewhere
[16].)  Here we have only scratched the surface and many interesting
solutions are still to be found. For example so far we have no
genuinely new solutions where the parameters are related in a
nonlinear way. What we now need is a fruitful ansatz, probably it will
come from physical applications.

\heading Acknowledgements \endheading
I would like to thank F. Nijhoff for introducing me to the subject and
the relevant references, and for discussions and comments on the
manuscript.

\Refs

\item{[1]} C.N. Yang, Phys. Rev. Lett. {\bf 29}, 1312 (1967);
R.J. Baxter, {\it Exactly Solved Models in Statistical Mechanics},
(Academic Press, 1982).

\item{[2]} P.P. Kulish and E.K. Sklyanin, "Quantum spectral transform
method. Recent developments", in {\it Integrable Quantum Field
Theories}, J. Hietarinta and C. Montonen (eds.) (Springer, 1982) pp.
61-119; L. Faddeev, in {\it Integrable systems, Nankai lectures 1987},
M.-L. Ge and X.C. Song (eds.) (World Scientific, 1989), pp. 23-70.

\item{[3]} M. Jimbo, "Introduction to the Yang-Baxter Equation", in [5];
M. Jimbo (ed.), {\it Yang-Baxter Equation in Integrable Systems},
World Scientific (1990).

\item{[4]} L.A. Takhtajan, "Lectures on Quantum Groups", in
{\it Introduction to Quantum Group and Integrable Massive Models of
Quantum Field Theory}, M.-L. Ge and B.-H. Zhao (eds.) (World
Scientific, 1990) pp. 69-197.

\item{[5]} C.N. Yang and M.L. Ge (eds.), {\it Braid Group, Knot
Theory and Statistical Mechanics} (World Scientific, 1989); L.H.
Kauffman, {\it Knots and Physics} (World Scientific, 1991).

\item{[6]} A.B. Zamolodchikov, Sov. Phys. JETP {\bf 52}, 325 (1981);
Comm. Math. Phys. {\bf 79}, 489 (1981).

\item{[7]} V.V. Bazhanov and Yu.G. Stroganov, Theor. Math. Phys.
{\bf 52}, 685 (1982); Nucl. Phys. {\bf B230}, 435 (1984).

\item{[8]} R.J. Baxter, Comm. Math. Phys. {\bf 88}, 185 (1983);
Phys. Rev. Lett. {\bf 53}, 1795 (1984); Physica {\bf 18D}, 321 (1986).

\item{[9]} J.M. Maillet and F. Nijhoff, Phys. Lett. A {\bf 134}, 221 (1989).

\item{[10]} I. Frenkel and G. Moore, Comm. Math. Phys. {\bf 138}, 259 (1991).

\item{[11]} J.S. Carter and M. Saito: "On Formulations and Solutions
of Simplex Equations", preprint.

\item{[12]} F. Nijhoff, private communication.

\item{[13]} J. Hietarinta, Phys. Lett. A {\bf 165}, 245 (1992).

\item{[14]} A.C. Hearn, {\it REDUCE User's Manual Version 3.4},
RAND Publication CP78 (7/91).

\item{[15]} H. Melenk, H.M. M\"oller and W. Neun, {\it GROEBNER,
A Package for Calculating Groebner Bases}. (Included in the
REDUCE 3.4 distribution package.)

\item{[16]} F. Nijhoff and J. Hietarinta, work in progress.

\enddocument

\bye